\documentclass[sigconf]{sig-alternate-per}

\usepackage{lineno, hyperref, amsmath, bbm, graphicx, caption, subcaption, float, mathtools, xcolor}

\DeclarePairedDelimiter{\floor}{\lfloor}{\rfloor}
\usepackage[font=small,skip=0pt]{caption}

\newtheorem{theorem}{Theorem}
\newtheorem{corollary}{Corollary}

\makeatletter
\g@addto@macro\normalsize{%
  \setlength\abovedisplayskip{2pt}
  \setlength\belowdisplayskip{2pt}
}
\makeatother

\def\sharedaffiliation{%
\end{tabular}
\begin{tabular}{c}}
\begin{document}
\title{Effective Straggler Mitigation: Which Clones Should Attack and When?}

\numberofauthors{3}
\author{
  Mehmet Fatih Akta\c{s} \\
  mehmet.aktas@rutgers.edu
  \alignauthor
  Pei Peng \\
  pei.peng@rutgers.edu
  \alignauthor
  Emina Soljanin \\
  emina.soljanin@rutgers.edu
  \sharedaffiliation
    \affaddr{Department of Electrical and Computer Engineering, Rutgers University}
}

\maketitle
\section{Introduction and Model}
\noindent\textbf{Motivation:}
Distributed (computing) systems aim to attain scalability through parallel execution of multiple tasks constituting a job. Each of these tasks is run on a separate node, and the job is completed only when the slowest task is finished. It has been observed that task execution times have significant variability, e.g., because of multiple job resource sharing \cite{dean2013tail}. The slowest tasks that determine the job execution time are known as "stragglers''.

Two common performance metrics for distributed job execution are 1) \emph{Latency,} measuring the execution time, and 2) \emph{Cost,} measuring the resource usage. Job execution is desired to be fast and with low cost, but these are conflicting objectives. Replicating tasks and running the replicas over separate nodes has been shown to be effective in mitigating the effect of stragglers on latency \cite{ananthanarayanan2013effective}, and is used in practice \cite{dean2008mapreduce}. Recent research proposes to delay replication, and clone only the tasks that at some point appear to be straggling, in order to reduce the cost \cite{wang2015using}.

Erasure coding is a more general form of redundancy than simple replication, and it has been considered for stragglers mitigation in both data download \cite{joshi2015queues} and, more recently, in distributed computing context \cite{dutta2016short}.
We here take this line of work further by analyzing the effect of coding on the tradeoff between latency and cost. As in \cite{wang2015using}, that deals with this issue in the context of replication, we consider systems where coded redundancy is introduced with a delay in order to reduce the cost, and examine the impact of that delay on latency. In \cite{ananthanarayanan2013effective}, introduction of redundancy has been playfully described as attack of the clones.
We here examine whether the redundancy should be simple replication or coding and when it should be introduced. That is, following the analogy of \cite{ananthanarayanan2013effective}, we ask which clones should attack and when.
\\[1ex]
\noindent\textbf{System Model:}
In our system, a job is split into $k$ tasks. The job execution starts with launching all its $k$ tasks, and the redundancy is introduced only if the job is not completed by some time $\Delta$.

\begin{figure}[t]
  \centering
  \includegraphics[width=0.5\textwidth, keepaspectratio=true]{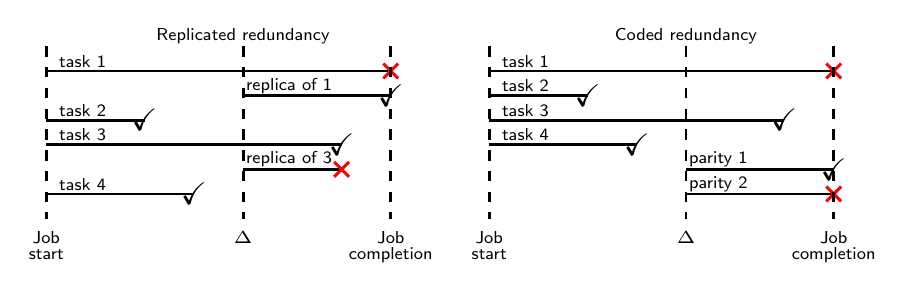}
  \caption{A job with four tasks is executed with delayed redundancy. Check mark represents completion of a task while cross represents cancellation of remaining outstanding redundant tasks.}
  \label{fig:fig_delayed_red}
\end{figure}

In replicated-redundancy $(k, c, \Delta)$-system, if the job still runs at time $\Delta$, $c$ replicas for each remaining task are launched. In coded-redundancy $(k, n, \Delta)$-system, if the job still runs at time $\Delta$, $n-k$ redundant parity tasks are launched where completion of any $k$ of all launched tasks results in total job completion (see Fig.~\ref{fig:fig_delayed_red}). Note that this assumption does not impose severe restrictions. Any linear computing algorithm can be structured in this way simply by using linear erasure codes. Particular examples can be found in e.g., \cite{dutta2016short} and references therein.

We assume that task execution times are iid and follow one of the three canonical distributions: 1) $Exp(\mu)$; commonly used to model execution of small-size tasks, 2) $SExp(D, \mu)$; constant $D$ plus $Exp(\mu)$ noise, used when the job size affects the execution time \cite{wang2015using}, (3) $\textit{Pareto}(\lambda, \alpha)$; canonical heavy-tail distribution that is observed to fit task execution times in real computing systems \cite{dean2013tail, reiss2012towards}.

We use $T$ to denote the job execution time. Cost is defined as the sum of the lifetimes of each task involved in job execution. There are two main setups that define cost: 1) Cost with task cancellation $C^c$; remaining outstanding tasks are canceled upon the job completion, which is a viable option for distributed computing with redundancy, 2) Cost without task cancellation $C$; tasks remaining after job completion run until they complete, which, for instance, is the only option for data transmission over multi-path network with redundancy.

In this paper, we analyze the effect of replicated and coded redundancy on cost and latency tradeoff. Specifically, we present exact expressions for expected latency and cost under delayed and zero-delay redundancy schemes. From these expressions, we observe that pain and gain of redundancy are strongly correlated with the tail of task execution time.
\\[1ex]
\textbf{Summary of Observations:} Coding allows us to increase degree of redundancy with finer steps than replication, which translates into greater achievable cost vs.\ latency region. Delaying coded-redundancy is not effective to trade off latency for cost, therefore, primarily the degree of redundancy should be tuned for the desired cost and latency. Coding is shown to outperform replication in terms of cost and latency together. When the task execution time has heavy tail, redundancy can reduce cost and latency simultaneously, where the reduction depends on how heavy the tail is.
\newpage
\section{Results and Observations}
We next state expressions for the expected latency and cost under replicated and coded redundancy. Note that these quantities depend on $k$, the number of tasks the job is split into, the redundancy level ($c$ in the replicated and $n$ in the coded systems), as well as $\Delta$, the time when the redundancy is introduced.
\\[1ex]
\noindent\textbf{Notation:}
$H_n$ is the $n$th harmonic number defined for $n \in Z^+$ as $\sum_{i=1}^n \frac{1}{i}$ and for $n \in R$ as $\int_0^1 \frac{1-x^n}{1-x} dx$. Incomplete Beta function $B(q;m,n)$ is defined for $q \in [0,1]$, $m, n \in R^+$ as $\int_0^q u^{m-1}(1-u)^{n-1} du$ and Beta function as $B(m,n) = B(1;m,n)$. Gamma function $\Gamma(x)$ is defined as $\int_0^{\infty} u^{x-1}e^{-u}du$ for $x \in R$ and as $(x-1)!$ for $x \in Z^+$.
\\[1ex]
\textbf{Expected Latency and Cost with Replication:}
\vspace{-1ex}
\begin{theorem}
 Under the exponential task execution time $X \sim Exp(\mu)$, expected latency in the replication $(k, c, \Delta)$-system is well approximated as
  \begin{equation*}
    E[T] \approx \frac{1}{\mu}(H_k - \frac{c}{c+1}H_{k-kq}).
  \label{eq:eq_reped_k_Exp__E_T}
  \end{equation*}
  Expected cost with ($C^c$) and without ($C$) task cancellation
  \begin{equation*}
    E[C^c] = \frac{k}{\mu}, \quad\quad E[C] = (c(1-q) + 1)\frac{k}{\mu}.
  \label{eq:eq_reped_k_Exp__E_C}
  \end{equation*}
  where $q = 1 - e^{-\mu\Delta}$.
  \label{thm_reped_k_Exp__E_T_E_C}
\end{theorem}

\begin{theorem}
  Under the shifted exponential task execution time $X \sim SExp(\frac{D}{k}, \mu)$, expected latency in the replication $(k, c, \Delta)$-system is well approximated as
  \begin{equation*}
    E[T] \approx \frac{D}{k} + \frac{1}{\mu}(H_k - \frac{c}{c+1}H_{k-kq}),
    ~\text{where $q = 1 - e^{-\mu\Delta}$}
  \label{eq:eq_coded_k_SExp__E_T}
  \end{equation*}
  
  Expected cost with ($C^c$) and without ($C$) task cancellation
  \begin{equation*}
  \begin{split}
    E[C^c] &= D + \frac{k}{\mu}(1 + c(1-q-e^{-\mu\Delta})), \quad \Delta > \frac{D}{k}, \\
    E[C] &= (c(1-q)+1)(D + \frac{k}{\mu}).
  \end{split}
  \label{cl_reped_k_SExp__E_C}
  \end{equation*}
   where $q = 1 - e^{-\mu(\Delta-\frac{D}{k})}$.
 \end{theorem}

\noindent
\textbf{Expected Latency and Cost with Coding:}
\vspace{-1ex}
\begin{theorem}
  Under the exponential task execution time $X \sim Exp(\mu)$, expected latency in coded redundancy $(k, n, \Delta)$-system is well approximated as
  \begin{equation*}
    E[T] \approx \Delta - \frac{1}{\mu}(B(q;k+1,0) + H_{n-kq} - H_{n-k}).
  \label{eq:eq_coded_k_n_Exp__E_T}
  \end{equation*}
  Expected cost with ($C^{c}$) and without ($C$) task cancellation
  \begin{equation*}
  \begin{split}
  	E[C^c] = \frac{k}{\mu}, \quad\quad E[C] = \frac{k}{\mu}q^k + \frac{n}{\mu}(1-q^k).
  \end{split}
  \label{eq:eq_coded_k_n_Exp__E_C}
  \end{equation*}
  where $q = 1 - e^{-\mu\Delta}$.
  \label{thm_coded_k_n_Exp__E_T_E_C}
\end{theorem}

\begin{theorem}
  Under the shifted exponential task execution time $X \sim SExp(\frac{D}{k}, \mu)$, expected latency in coded redundancy $(k, n, \Delta)$-system is well approximated as
  \begin{equation*}
  \begin{split}
    E[T] \approx \frac{D}{k} + \Delta - \frac{1}{\mu}(B(q;k+1,0) + H_{n-kq} - H_{n-k}).
  \end{split}
  \label{eq:eq_coded_k_n_SExp__E_T}
  \end{equation*}
  where $q = 1 - e^{-\mu\Delta}$.
  
  Expected cost with ($C^{c}$) and without ($C$) task cancellation
  \begin{equation*}
  \begin{split}
	E[C] &= q^k k\left(\frac{1}{\mu}+\frac{D}{k} \right) + (1-q^k)n\left(\frac{1}{\mu}+\frac{D}{k} \right), \\
    E[C^c] &\approx E[C] - \frac{(n-k)}{\mu}(1-q^k) \\
    &- \frac{(n-k)}{\mu}\eta^{-k(1-q)} B(\eta;k-kq+1,0)(\tilde{q}^k-q^k).
  \end{split}
  \label{eq:eq_coded_k_n_SExp__E_C}
  \end{equation*}
  where $q = \mathbbm{1}(\Delta > \frac{D}{k})(1-e^{-\mu(\Delta-\frac{D}{k})})$, $\tilde{q} = 1-e^{-\mu\Delta}$ and $\eta = 1-e^{-\mu\Delta}$.
\end{theorem}

\noindent\textbf{Scheme Comparison:}
In order to answer the title question {\it which clones to send and when,}
we next compare replicated and coded redundancy in distributed computing context, where it is feasible to cancel the running redundant tasks upon the job completion.

With exponential task execution time, under both replicated and coded redundancy, the expected cost depends neither on the time $\Delta$ redundancy is introduced nor on the degree of redundancy $c$ and $n$ (see Thm~\ref{thm_reped_k_Exp__E_T_E_C} and \ref{thm_coded_k_n_Exp__E_T_E_C}). Consequently, in order to achieve the minimum latency, one can introduce all available redundancy at once ($\Delta=0$) with zero expected penalty in cost.

\begin{figure}[h]
  \centering
  \includegraphics[width=0.48\textwidth, keepaspectratio=true]{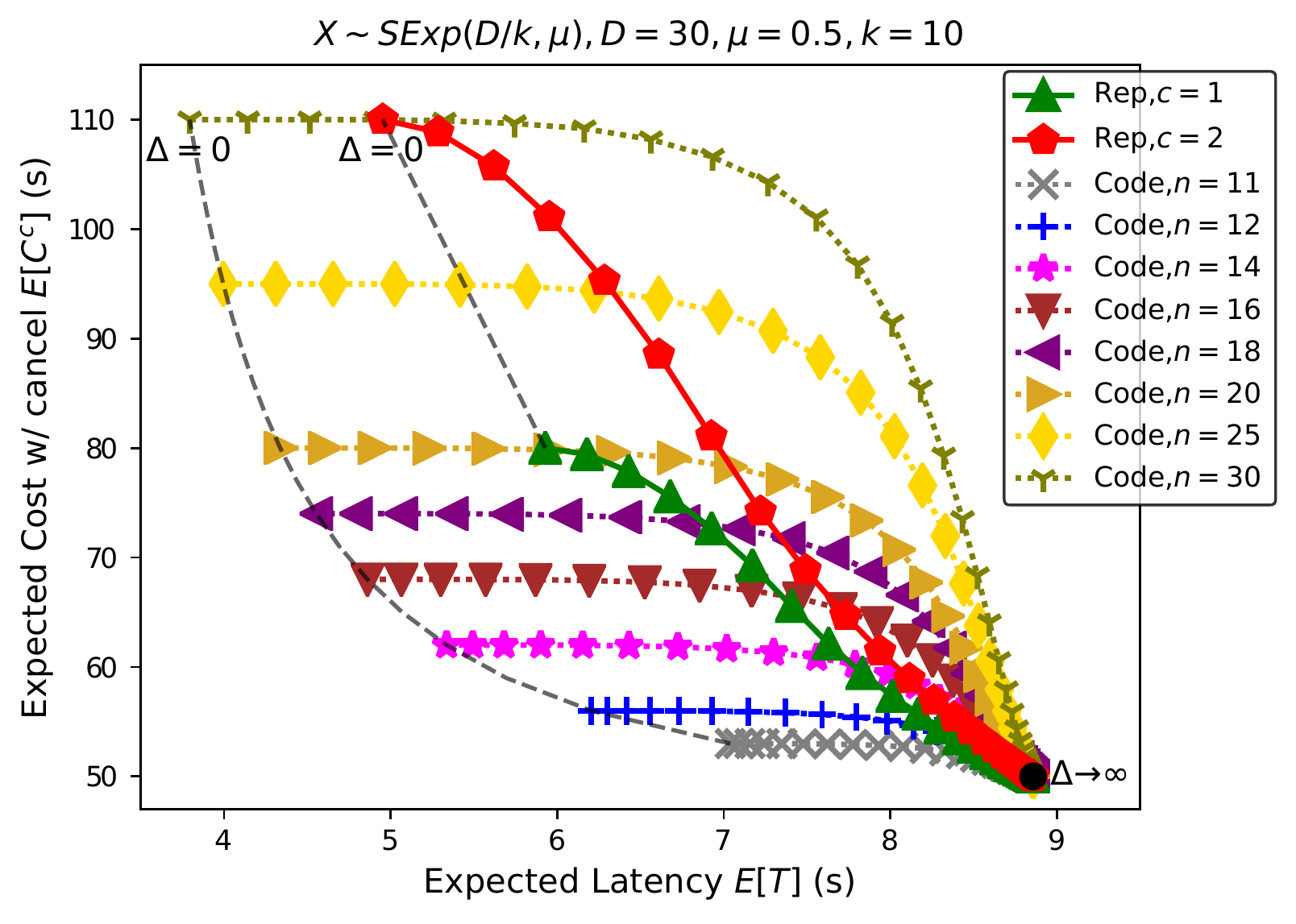}
  \caption{Under $\textit{SExp}$ task execution time, achievable expected cost with task cancellation vs. latency region is plotted for replicated ($c=1,2$) and coded ($n \in [k+1, 3k]$) redundancy by varying the time ($\Delta$) of introducing redundancy along each curve.}
  \label{fig:plot_reped_vs_coded_k_10}
\end{figure}

We want to understand the reduction in cost (gain) and increase in latency (pain) per increase in $\Delta$. Fig.~\ref{fig:plot_reped_vs_coded_k_10} shows cost vs.\ latency under delayed redundancy for $SExp$ tasks. For coded redundancy, we observe two phases: 1) Initially, increasing $\Delta$ away from $0$ returns almost no reduction in cost but significantly increases latency. 2) Beyond a certain point, increasing $\Delta$ further reduces cost significantly while not increasing delay much. In other words, significant reduction in cost by delaying redundancy is possible only with significant increase in latency. Therefore, delaying coded redundancy is not effective because one can simply achieve less cost for the same latency by decreasing degree of redundancy $n$. Simulations show that this two-phase behavior exists for $\textit{Pareto}$ task execution time as well. Note that delaying is effective for replicated redundancy to reduce cost up to some point, beyond which, once again, it is better to reduce the degree of replication $c$.

\begin{figure*}[hbt]
  \centering
  \begin{subfigure}[]{.32\textwidth}
    \centering
    \includegraphics[width=1\textwidth, keepaspectratio=true]{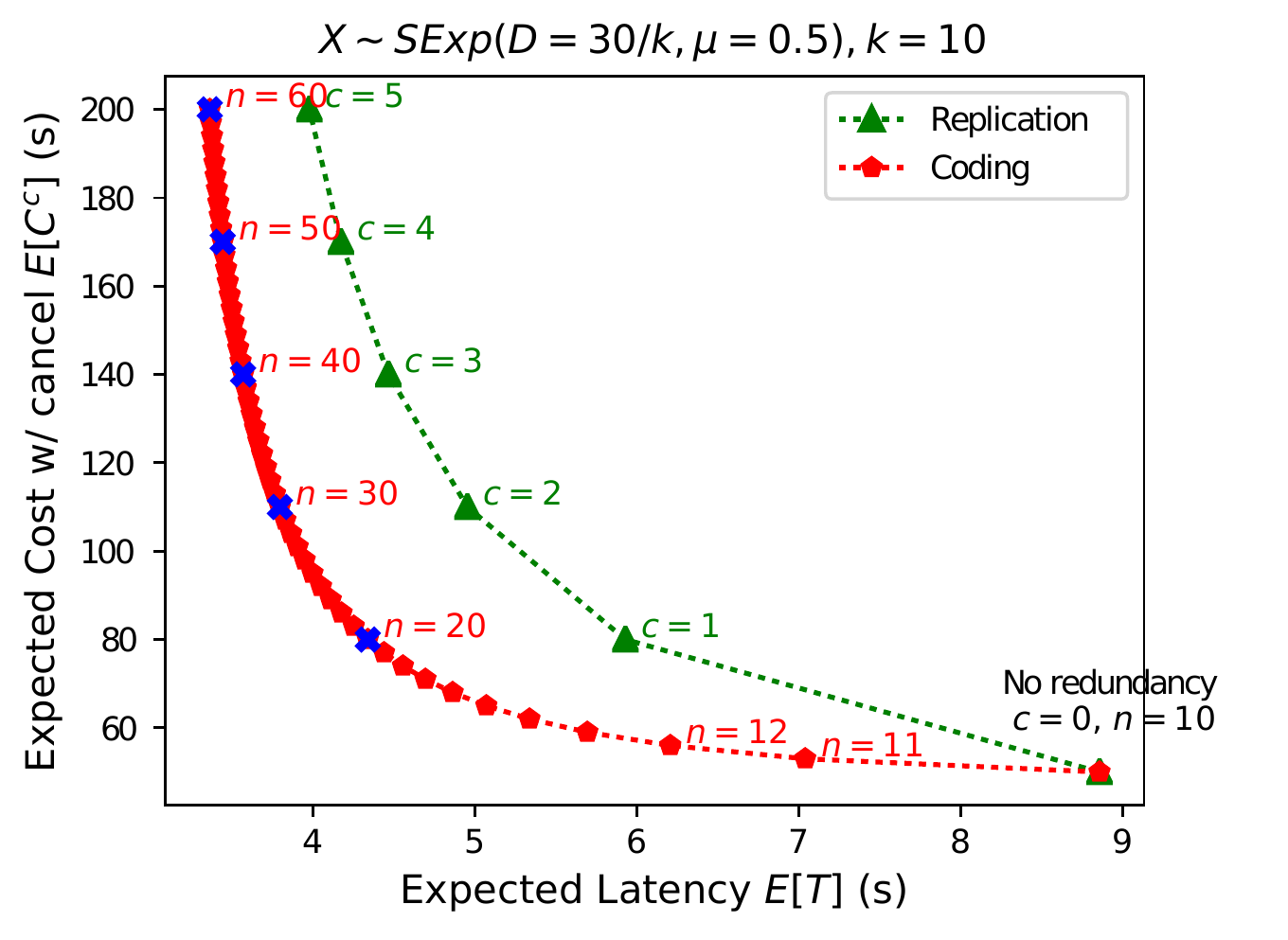}
  \end{subfigure}
  \begin{subfigure}[]{.32\textwidth}
    \centering
    \includegraphics[width=1\textwidth, keepaspectratio=true]{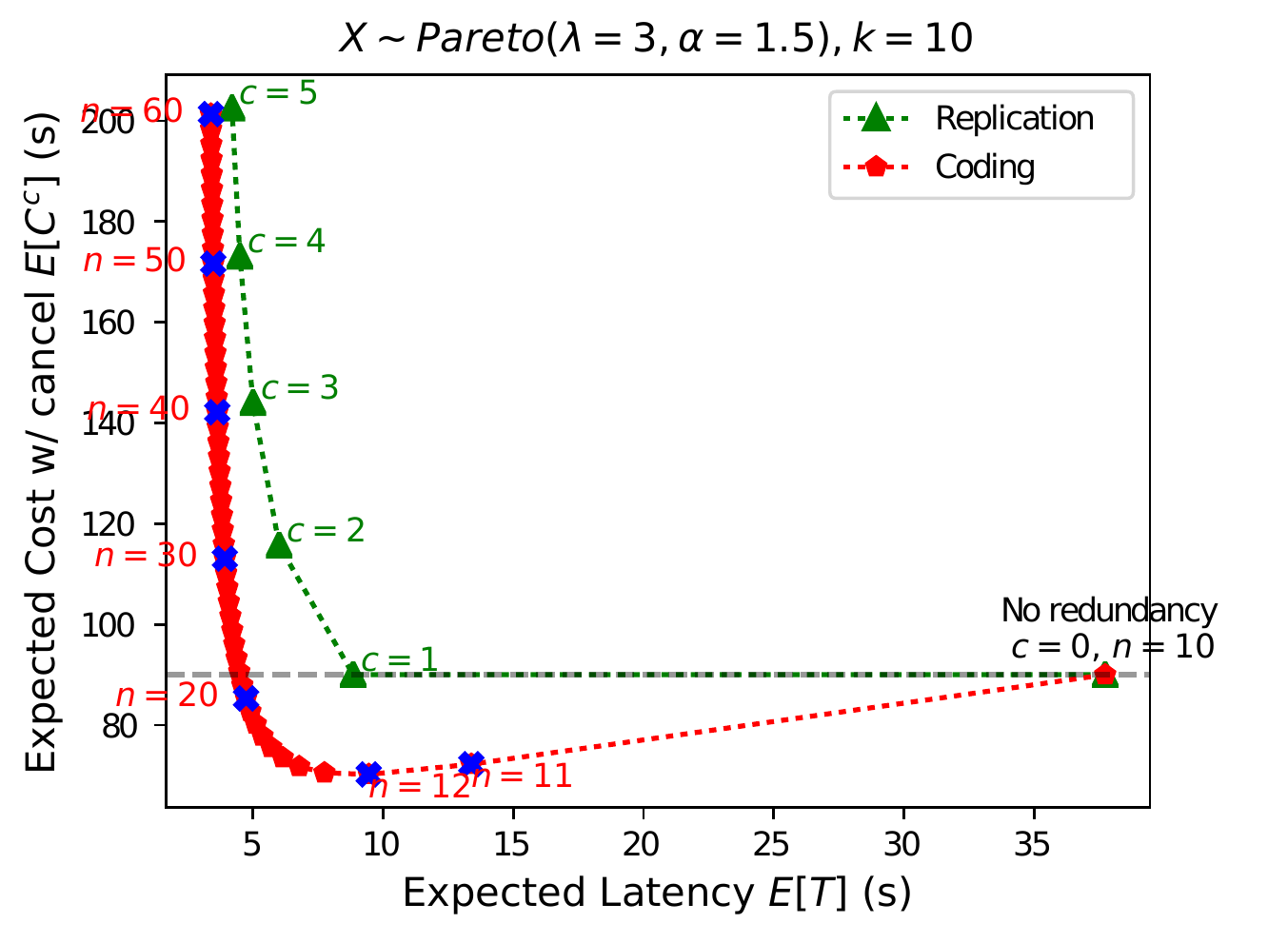}
  \end{subfigure}
  \begin{subfigure}[]{.32\textwidth}
    \centering
    \includegraphics[width=1\textwidth, keepaspectratio=true]{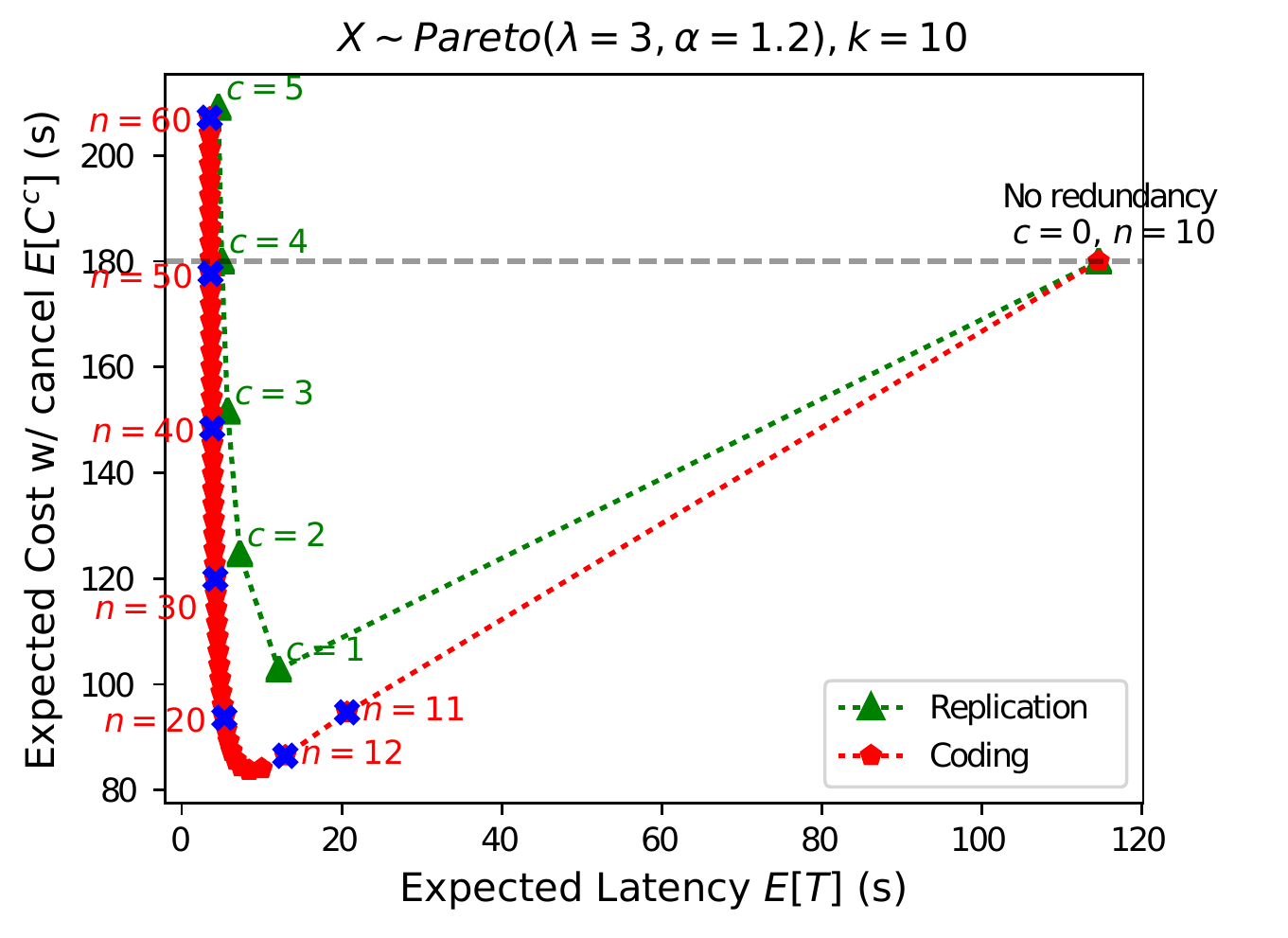}
  \end{subfigure}
  \caption{Expected cost vs.\ latency for zero-delay redundancy where redundancy levels $c$ and $n$ vary along the curves. Tail heaviness increases from left to right. The heavier the tail is, the higher the maximum reduction in expected cost and latency is.}
  \label{fig:plot_zerodelay_reped_vs_coded_k_10}
\end{figure*}
 Thm.~\ref{theo_zero_delay_red_E_T__E_C} gives exact expressions for the expected cost and latency under zero-delay redundancy. Under both $SExp$ and $Pareto$ task execution time, coding always achieves better expected cost and latency than replication as illustrated in Fig.~\ref{fig:plot_zerodelay_reped_vs_coded_k_10}.
\begin{theorem}
  Let expected latency and cost with task cancellation be $E[T_{(k,c)}]$, $E[C_{(k,c)}]$ for zero-delay replicated redundancy, and $E[T_{(k,n)}]$, $E[C_{(k,n)}]$ for zero-delay coded redundancy. Under task execution time $X \sim SExp(\frac{D}{k}, \mu)$,
  \begin{equation*}
  \begin{split}
    E[T_{(k,c)}] &= \frac{D}{k} + \frac{H_k}{(c+1)\mu}, \quad\quad E[C_{(k,c)}] = (c+1)D + \frac{k}{\mu}, \\
    E[T_{(k,n)}] &= \frac{D}{k} + \frac{1}{\mu}(H_n-H_{n-k}), \quad E[C_{(k,n)}] = \frac{nD}{k} + \frac{k}{\mu}.
  \end{split}
  \label{eq:eq_zerodelay_reped_SExp_E_T__E_C}
  \end{equation*}
  
  Under task execution time $X \sim Pareto(\lambda, \alpha)$,
  \begin{equation*}
  \begin{split}
    E[T_{(k,c)}] &= \lambda k!\frac{\Gamma(1-((c+1)\alpha)^{-1})}{\Gamma(k+1-((c+1)\alpha)^{-1})}, \\
    E[C_{(k,c)}] &= \lambda k(c+1)\frac{(c+1)\alpha}{(c+1)\alpha-1}, \\
    E[T_{(k,n)}] &= \lambda\frac{n!}{(n-k)!}\frac{\Gamma(n-k+1-\alpha^{-1})}{\Gamma(n+1-\alpha^{-1})}, \\
    E[C_{(k,n)}] &= \lambda\frac{n}{\alpha-1}(\alpha - \frac{\Gamma(n)}{\Gamma(n-k)}\frac{\Gamma(n-k+1-\alpha^{-1})}{\Gamma(n+1-\alpha^{-1})}).
  \end{split}
  \label{eq:eq_zerodelay_reped_Pareto_E_T__E_C}
  \end{equation*}
  \label{theo_zero_delay_red_E_T__E_C}
\end{theorem}

One would expect that adding more redundancy reduces latency but always increases cost. In \cite{wang2015using} replicated redundancy is demonstrated to reduce both cost and latency under heavy-tail task execution time. Fig.~\ref{fig:plot_zerodelay_reped_vs_coded_k_10} shows and compares this for replicated and also the coded redundancy using the analytical expressions presented here. Under heavy-tail, it is possible to reduce latency by adding redundancy and still pay for the baseline cost of running without redundancy. Corollary \ref{cor_zerodelay_red_pareto_reduc_in_E_T_for_baseline_E_C} gives expressions for the minimum achievable expected latency without exceeding the baseline cost.
\begin{corollary}
  Under task execution time $X \sim \textit{Pareto}(\lambda, \alpha)$ in zero-delay replicated redundancy system, minimum latency $E[T_{\min}]$ that can be achieved without exceeding the baseline cost is,
  \begin{equation*}
  \begin{split}
    E[T_{\min}] = \lambda k!\frac{\Gamma(1-(\alpha(c_{\max}+1))^{-1})}{\Gamma(k+1-(\alpha(c_{\max}+1))^{-1})}.
  \end{split}
  \label{eq:eq_reped_pareto_reduc_in_E_T_for_same_E_C}
  \end{equation*}
  where $c_{\max} = \max\{\floor*{\frac{1}{\alpha-1}}-1, 0\}$ and any reduction in latency without exceeding the baseline cost is possible only if $\alpha < 1.5$.
  For coded redundancy system, a tight upper bound on $E[T_{\min}]$ is
  \begin{equation*}
  \begin{split}
    E[T_{\min}] < \lambda\alpha + \lambda k!\frac{\Gamma(1-\alpha^{-1})}{\Gamma(k+1-\alpha^{-1})}.
  \end{split}
  \label{eq:eq_zerodelay_coded_pareto_reduc_in_E_T_for_same_E_C}
  \end{equation*}
\label{cor_zerodelay_red_pareto_reduc_in_E_T_for_baseline_E_C}
  
\end{corollary}

\begin{figure}[h]
  \centering
  \includegraphics[width=0.35\textwidth, keepaspectratio=true]{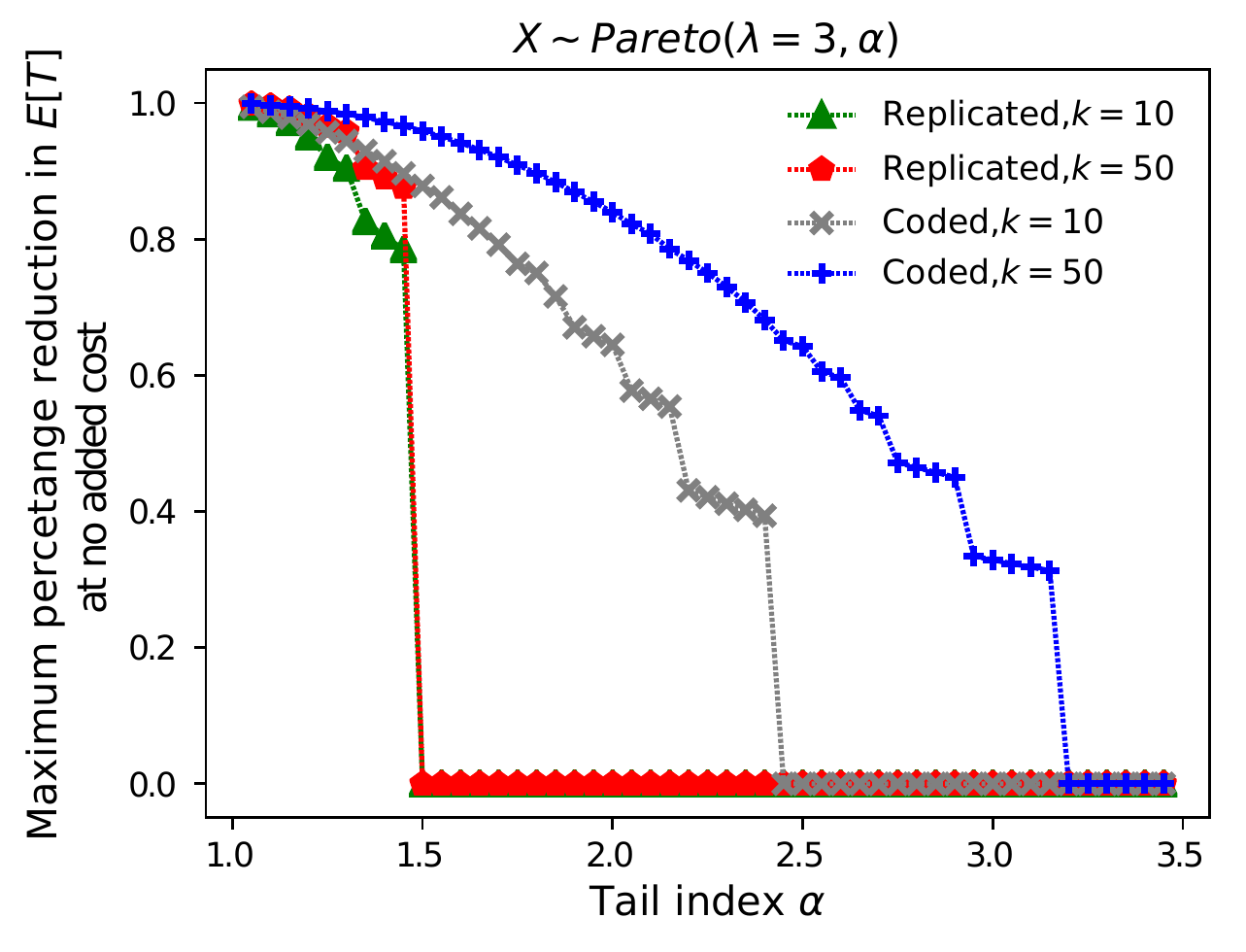}
  \caption{$\frac{E[T_0]-E[T_{\min}]}{E[T_0]}$ vs. $\alpha$. $E[T_{\min}]$ is the minimum expected latency with redundancy without exceeding the baseline cost and $E[T_0]$ is the expected latency with no redundancy.}
  \label{fig:plot_zerodelay_red_pareto_reduc_in_E_T_for_baseline_E_C}
\end{figure}

Fig.~\ref{fig:plot_zerodelay_red_pareto_reduc_in_E_T_for_baseline_E_C} illustrates that the maximum percentage reduction in latency while paying for less than the baseline cost depends on the tail of task execution time. As stated in Corollary \ref{cor_zerodelay_red_pareto_reduc_in_E_T_for_baseline_E_C}, this reduction is possible under replicated redundancy only when the tail index is less than $1.5$, in other words when the tail is very heavy, while coding relaxes this constraint significantly. In addition, the constraint on $\alpha$ is independent of the number of tasks $k$ under replication, while it increases with $k$ under coding, meaning that jobs with larger number of tasks can get reduction in latency at no cost even for lighter tailed task execution times.

\bibliographystyle{unsrt}
\bibliography{references}

\end{document}